\hd{3. A binary tree with transversely 2-periodic potentials}

We now specialize to a binary tree. 

\vskip 20pt

\epsfxsize=4cm
\centerline{\epsffile{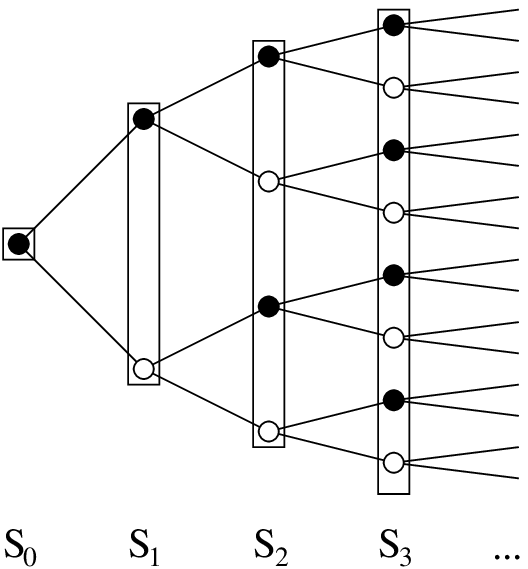}}

For a tree, the forward Green's functions are diagonal, and with
$$\eqalign{
G_{n+1}(\lambda) &= \diag[z_1,z_2, \ldots, z_{2^{n+1}}],\cr
Q_n &= \diag[q_1,q_2, \ldots, q_{2^{n}}],
}$$
we have
$$
\Phi_n(G_{n+1},Q_n,\lambda) = \diag\left[{{-1}\over{z_1+z_2+\lambda-q_1}}, \ldots, {{-1}\over{z_{2^{n+1}-1}+z_{2^{n+1}}+\lambda-q_{2^n}}}\right].
$$

To define a two-periodic potential we choose for each sphere (except the root)
two potential values $\qq = (q_1, q_2)$ at random, independently for each sphere, 
according to an identical joint distribution 
$\nu$. In the  diagram, the spheres are outlined by boxes. For each sphere (except the first), after choosing $\qq=(q_1,q_2)$, we set the potential at all the black vertices equal to $q_1$ and
the potential at all the white vertices equal to $q_2$. The potential value at $0$ is chosen according
to some single site distribution $\nu_0$.

We make the following assumptions about this distribution $\nu$. The distribution has bounded support:
\be{nuassn1}
\hbox{$\nu$ is supported in $\{\qq=(q_1,q_2) : |q_1|\le 1, |q_2|\le 1\}$}.
\ee
The distribution is centred on zero:
\be{nuassn3}
\int_{\RR^2} (q_1 + q_2) d\nu(\qq) = 0.
\ee
Let $c_{ij} = \int_{\RR^2} q_iq_j d\nu(\qq)$. Then
\be{nuassn2}
c=c_{11}+c_{22} >0 \quad\hbox{and}\quad \delta = {{2c_{12}}\over{c_{11}+c_{22}}} < {{1}/{2}}.
\ee
The first inequality in \rf{nuassn2} simply says that $q$ is not identically zero. The second is a bound on the correlation.
Completely correlated potentials (that is, the one-dimensional case where the spectrum is localized) would correspond to 
$\delta = 1$.

To adjust the disorder, we multiply the potential by a coupling constant $a>0$ and study the Schr\"odinger operator
$H_a=L+a\,Q$. This amounts to replacing $\nu$ with
the scaled distribution $\nu_a$ satisfying               
$$
\int_{\RR^2} f(\qq)\, d\nu_a(\qq) = \int_{\RR^2} f(a\qq)\, d\nu(\qq).
$$
The scaled distribution $\nu_a$ is supported
in $\{\qq=(q_1,q_2) : |q_1|\le a, |q_2|\le a\}$.

We can now formulate the main theorem for this section.
\begintheoremlabel{twoperiodic}
Let $\nu$ be a probability measure on $\RR^2$ satisfying \rf{nuassn1}, \rf{nuassn3} and \rf{nuassn2} and
let $H_a$ be the random discrete Schr\"odinger operator on the binary tree corresponding to the transversely two-periodic potential defined by the scaled distribution $\nu_a$. There exists $\lambda_0 \in (0,2\sqrt{2})$ such that for sufficiently small $a$ the spectral measure for $H_a$ corresponding to $\delta_0$ has purely absolutely continuous spectrum in $(-\lambda_0,\lambda_0)$.
\endtheoremlabel

For a two-periodic potential, formula \rf{probrecursion} can be simplified.
In this case the measure $N_n$ is independent of $n$ and
concentrated on the two-dimensional hyperplane
where $q_1=q_3=q_5=\cdots$ and $q_2=q_4=q_6=\cdots$. Thus, introducing a coupling constant $a$, the measure $N_n$
is a product of $\nu_a$ with delta functions for the hyperplane. 
For these potentials the diagonal entries of
$G_n(\lambda)$ exhibit the same symmetry as the potentials, so the probability distribution for $G_n(\lambda)$
is determined by the joint distribution $r_{a,\lambda}$ for $(z_1, z_2)$, which also is independent of $n$. 
With this notation, formula
\rf{probrecursion} can be written

$$\displaylines{\quad\quad
\int_{\HH\times\HH} f(z_1,z_2)\, dr_{a,\lambda}(z_1,z_2)
\hfill\cr\hfill
=\int_{\HH\times\HH\times\RR^2} f\left(-{{1}\over{z_1+z_2+\lambda-q_1}},-{{1}\over{z_1+z_2+\lambda-q_2}}\right)\,d\nu_a(\qq)\,dr_{a,\lambda}(z_1,z_2).
\quad\quad\cr}$$

It is convenient to introduce a new random variable $u=z_1+z_2+\lambda$ for every sphere except the first.
Let $\rho_{a,\lambda}$ denote the distribution on 
$\HH$ for $u$. Then, taking $f(z_1,z_2)=g(z_1+z_2+\lambda)$ in the formula above we obtain our main recursion formula
\be{grecur}
\int_\HH g(u)\, d\rho_{a,\lambda}(u) = \int_{\HH\times\RR^2} g(\phi_{\qq,\lambda}(u))\, d\nu_a(\qq)\, d\rho_{a,\lambda}(u),
\ee
where
\be{phidef}
\phi_{\qq,\lambda}(u) =  -{{1}\over{u-q_1}} - {{1}\over{u-q_2}} + \lambda.
\ee
A source of difficulty is the singular behaviour of $\phi_{\qq,\lambda}$ near the diagonal of $\qq$. When $q_1=q_2$
(and $\Im(\lambda)\ge 0$),
then $\phi_{\qq,\lambda}$ is a linear fractional transformation that defines an injective map from $\HH$ to $\HH$. In fact, if $\lambda\in\RR$ the map is a hyperbolic isometry. However, as soon as $q_1\ne q_2$ the map $\phi_{\qq,\lambda}$
covers $\HH$ twice. This can be seen even when we only consider real values of $u$. In this case $\phi_{\qq,\lambda}(u)$ ranges over all of $\RR$ for $u$ in the interval $(q_1,q_2)$ (supposing for the moment that $q_1 < q_2$). This interval shrinks and then disappears as $q_1$ approaches $q_2$.

We now introduce the weight function $\cd$. For $\lambda\in(-2\sqrt{2}, 2\sqrt{2})$ the fixed point solution of $u\mapsto \phi_{{\bf 0},\lambda}(u)$ is $u_\lambda = \lambda/2 + i\sqrt{2-\lambda^2/4}$. Define
\be{cddef}
\cd(u) = {{|u-u_\lambda|^2}\over{\Im (u)}}.
\ee

Our goal is to bound the moment
\be{Mdef}
M_{a,\alpha,\lambda} = \int_\HH \cd(u)^{1+\alpha} d\rho_{a,\lambda}(u).
\ee
Given \thmrf{KS}, such a bound for $R_{0,\lambda}$ in place of $\rho_{a,\lambda}$
will provide a proof of \thmrf{twoperiodic}. This is done in the following lemma.
\beginlemmalabel{Mimplies}
Suppose that 
$$
\sup_{{|\Re\lambda|\le\lambda_0}\atop{0<\Im\lambda\le\epsilon}} M_{a,\alpha,\lambda} < C
$$
for some positive $a$, $\alpha$ and $\epsilon$. Then the spectral measure for $\delta_0$ corresponding to
the transversely two-periodic random
potential with coupling constant $a$ has purely absolutely continuous spectrum in $[-\lambda_0,\lambda_0]$.
\endlemmalabel
\beginproof
Let $w(z) = |z-i|^2/\Im(z)$. The recursion formula \rf{probrecursion} for the first level implies
$$\eqalign{
\int_\HH w(z)^{1+\alpha}\, dR_{0,\lambda}(z) &= \int_{\HH\times\RR} w(-(u-q)^{-1})^{1+\alpha}\, d\nu_0(q)\, d\rho_{a,\lambda}(u)\cr
&\le \left(\sup_{u\in\HH} \int_\RR \left({{w(-(u-q)^{-1})}\over{\cd(u)+1}}\right)^{1+\alpha}\, d\nu_0(q)\right)(M_{a,\alpha,\lambda}+1),\cr
}$$
so the lemma follows from \thmrf{KS} and the bound
$$
\sup_{u\in\HH} \int_\RR \left({{w(-(u-q)^{-1})}\over{\cd(u)+1}}\right)^{1+\alpha}\, d\nu_0(q) \le C.
$$
This is true, for example, for any distribution $\nu_0$ with bounded support.
\endproof

\hd{4. Bounding $M_{a,\alpha,\lambda}$}

\thmrf{Mimplies} shows that our main theorem follows from a bound for $M_{a,\alpha,\lambda}$.
We now explain how we can obtain such a bound.
Beginning with \rf{Mdef} we introduce a cutoff function $\chi$ with support in a neighbourhood of the boundary at infinity. Since $\cd$ is bounded on the compact support of $1-\chi$,
$$
M_{a,\alpha,\lambda} \le \int_\HH \chi(u)\;\cd(u)^{1+\alpha} d\rho_{a,\lambda}(u) + C,
$$
where $C$ only depends on the support of $\chi$. Now we apply the recursion formula \rf{grecur} to conclude
$$
M_{a,\alpha,\lambda} \le \int_\HH\int_{\RR^2} \chi(\phi_{\qq,\lambda}(u))\;\cd(\phi_{\qq,\lambda}(u))^{1+\alpha}\;  d\nu_a(\qq) d\rho_{a,\lambda}(u) + C.
$$
Since the image of $\phi_{\qq,\lambda}(u)$  is
compact, as $\qq$ ranges over the support of $\nu_a$, $\lambda$ ranges over the
rectangle $|\Re(\lambda)| \le \lambda_0$, $0\le\Im(\lambda)\le\epsilon$ and $u$ ranges over the support of $1-\chi$, the function $\cd(\phi_{\qq,\lambda}(\cdot))$ is bounded there, and we may again insert a cutoff and conclude
\be{M1}
M_{a,\alpha,\lambda} \le \int_\HH\int_{\RR^2} \chi(u)\; \chi(\phi_{\qq,\lambda}(u))\; \cd(\phi_{\qq,\lambda}(u))^{1+\alpha}\; d\nu_a(\qq) d\rho_{a,\lambda}(u) + C.
\ee
The constant $C$ is different from the previous equation, but can still be taken to be independent of
$\lambda$ in the range of values we are considering. 

Here is the essential idea of our argument. Introduce 
\be{mudef}
\mu^0_{\qq,\lambda}(u) = {{\cd(\phi_{\qq,\lambda}(u))}\over{\cd(u)}}.
\ee
A calculation using the fact that $u_\lambda$ is the fixed
point for $\phi_{{\bf 0},\lambda}$ yields
$$
\mu^0_{\qq,\lambda}(u) = {{|(2u-q_1-q_2)u_\lambda - 2(u-q_1)(u-q_2)|^2}\Im(u)
\over{|u_\lambda|^2(\Im(u)(|u-q_1|^2 + |u-q_2|^2) + \Im(\lambda)|u-q_1|^2|u-q_2|^2)|u-u_\lambda|^2}}.
$$
To simplify the calculations we will actually work with the upper bound obtained by setting $\Im(\lambda)$ in the denominator to zero. Define
$$
\mu_{\qq,\lambda}(u) = {{|(2u-q_1-q_2)u_\lambda - 2(u-q_1)(u-q_2)|^2}\over{|u_\lambda|^2(|u-q_1|^2 + |u-q_2|^2)|u-u_\lambda|^2}}.
$$ 
Then $\mu^0_{\qq,\lambda} \le \mu_{\qq,\lambda}$. Now introduce the averaged version
$$
\overline\mu_{a,\alpha,\lambda}(u) = \int_{\RR^2} \mu_{\qq,\lambda}^{1+\alpha}(u)\; d\nu_a(\qq).
$$
Then \rf{M1} implies
$$
M_{a,\alpha,\lambda} \le \int_\HH \chi(u)\, \overline\mu_{a,\alpha,\lambda}(u)\; \cd(u)^{1+\alpha}d\rho_{a,\lambda}(u) + C.
$$
So if we knew that $\overline\mu_{a,\alpha,\lambda}(u) \le 1-\epsilon_1$ on $\supp(\chi)$ for a suitable range of $\lambda$, then we would obtain $M_{a,\alpha,\lambda}\le (1-\epsilon_1) M_{a,\alpha,\lambda} +C$ which gives the desired bound on $M_{a,\alpha,\lambda}$.

Averaging over $\qq$ will be essential for obtaining such a bound, since $\mu_{{\bf 0},\lambda}(u) = 1$ when $\lambda$ is real. 
Notice that $\mu_{\qq,\lambda}(u)$ is continuous as $u$ and $\lambda$ approach the real axis, except at $u=q_1=q_2$.
This includes $u=i\infty$, by which we mean continuity as $w\rarr 0$ when we set $u=-1/w$.
At the singular point we can define $\mu_{\qq,\lambda}(u)$ to be the supremum of all possible limits. In this way we
can extend $\mu_{\qq,\lambda}(u)$ to an upper semi-continuous function whose domain includes real values of $u$ and $\lambda$.

Here is the bound for $\overline\mu_{a,\alpha,\lambda}(u)$. This is the main technical result in the first part of the paper. 
\beginlemmalabel{mubarbdd} Suppose that $\nu$ is a probability measure on $\RR^2$ satisfying \rf{nuassn1}, \rf{nuassn3} and \rf{nuassn2}. Assume $u$ and $\lambda$ are real, $|\lambda|< 2\sqrt{2}$ and $a$ and $R$ are positive real numbers satisfying $R\ge 2$ and $aR\le 1/4$.
Then there exist positive constants $C_i$ such that with $c$ and $\delta$ defined by \rf{nuassn2}
\vskip 4pt
\be{mubarbdd}
\overline\mu_{a,\alpha,\lambda}(u) \le
\cases{
1 - \displaystyle{{{c(1-\delta)}\over{20 R^2}} + C_1 a R + C_2} \alpha & {\rm for} $|u| \le aR$ 
\cr
1 - \displaystyle{{{a^2 c}\over{|u|^2}}\left({{p(u,\lambda,\delta)}\over{2|u-u_\lambda|^2}}
-\left({{C_3}\over{R}} + C_4\alpha\right)\right)} & {\rm for} $|u| \ge aR$
\boxhwt{25pt}{0pt}{0pt}\cr
},
\ee
\vskip 7pt
\noindent where
$$
p(u,\lambda,\delta) = (1-2\delta)u^2-(1-\delta)\lambda u + 1-\delta.
$$
\endlemmalabel

This lemma is proved in a separate section.
When $|u|\rarr\infty$ the bound tends to $1$, so this bound alone is not sufficient. To procede we 
must iterate the procedure leading to \rf{M1}. Starting with \rf{M1} (with $\qq$ replaced by $\qq_1$) we
apply \rf{grecur} to arrive at
\be{Mbdd2}\eqalign{
M_{a,\alpha,\lambda} &\le \int_\HH\int_{\RR^2}\int_{\RR^2}\chi(u)\; \chi(\phi_{\qq_1,\lambda}(u))\;
\cd(\phi_{\qq_1,\lambda}\circ \phi_{\qq_2,\lambda}(u))^{1+\alpha}\;
d\nu_a(\qq_1)d\nu_a(\qq_2)\;d\rho_{a,\lambda}(u) + C\cr
&= \int_\HH
\int_{\RR^2}\int_{\RR^2}
\chi(u) \mu_{\qq_1,\lambda}^{1+\alpha}(u)
\;
\chi(\phi_{\qq_1,\lambda}(u)) \mu_{\qq_2,\lambda}^{1+\alpha}(\phi_{\qq_1,\lambda}(u))
d\nu_a(\qq_1)d\nu_a(\qq_2)
d\rho_{a,\lambda}(u) + C.\cr
}\ee
We used $\chi \le 1$ to drop one term involving $\chi$.

In view of \thmrf{Mimplies}, the following lemma will complete the proof of \thmrf{twoperiodic}.

\beginlemma
Suppose that $\nu$ satisfies \rf{nuassn1}, \rf{nuassn3} and \rf{nuassn2}. Then there exists $\lambda_0\in(0,2\sqrt{2})$ such that for small enough $a$ and $\epsilon$
$$
\sup_{{|\Re\lambda|\le\lambda_0}\atop{0<\Im\lambda\le\epsilon}} M_{a,\alpha,\lambda} < C.
$$
\endlemma

\beginproof
Let 
$$
m_{a,\alpha,\lambda}(u) = 
\chi(u)\int_{\RR^2}\int_{\RR^2}
 \mu_{\qq_1,\lambda}^{1+\alpha}(u)
\;
\chi(\phi_{\qq_1,\lambda}(u)) \mu_{\qq_2,\lambda}^{1+\alpha}(\phi_{\qq_1,\lambda}(u))
d\nu_a(\qq_1)d\nu_a(\qq_2).
$$
Given \rf{Mbdd2}, it suffices to show that there exists $\lambda_0\in(0,2\sqrt{2})$  and $\epsilon_1 >0$ so
that
\be{mgoal}
m_{a,\alpha,\lambda}(u) \le 1-\epsilon_1
\ee
for all $\lambda$ with $|\Re\lambda|\le\lambda_0$, $0\le\Im\lambda\le\epsilon$ and with $\epsilon$, $\supp(\chi)$, $a$ and $\alpha$ sufficiently small.
An obvious estimate for $m_{a,\alpha,\lambda}(u)$ is
\be{obv}
m_{a,\alpha,\lambda}(u) \le \chi(u)\, \overline\mu_{a,\alpha,\lambda}(u) 
\sup_{\qq_1\in\supp(\nu_a)}\Big[\chi(\phi_{\qq_1,\lambda}(u))\,\overline\mu_{a,\alpha,\lambda}(\phi_{\qq_1,\lambda}(u)) \Big].
\ee

We begin by choosing $\lambda_0$ with
$$
\lambda_0 < 2\sqrt{{{1-2\delta}\over{1-\delta}}}.
$$
Then a simple calculation shows that the polynomial $p(u,\lambda,\delta)$ in \thmrf{mubarbdd} is bounded below
$$
p(u,\lambda,\delta)\ge p_0>0
$$
for all $u,\lambda\in\RR$ with $|\lambda|\le\lambda_0$. Choosing $R$ sufficiently large and $\alpha$ suffciently small we can simplify the estimate in \thmrf{mubarbdd} to read
\vskip 4pt
$$
\overline\mu_{a,\alpha,\lambda}(u) \le
\cases{
1 - 2\epsilon_2  + C_1 a R & {\rm for} $|u| \le aR$ 
\cr
1 - \displaystyle{{{a^2 \epsilon_3}\over{|u|^2}}} & {\rm for} $|u| \ge aR$ 
\boxhwt{25pt}{0pt}{0pt}\cr
}
$$
\vskip 7pt
\noindent for some $\epsilon_2,\epsilon_3>0$ and for all $u\in \PA_\infty\HH = \RR \cup \{i\infty\}$
and real $\lambda$ with $|\lambda|\le\lambda_0$.
Then, choosing $a$ small (depending on $R$) we obtain
\vskip 4pt
\be{mubarbdd2}
\overline\mu_{a,\alpha,\lambda}(u) \le
\cases{
1 - \epsilon_2  & {\rm for} $|u| \le aR$ 
\cr
1 - \displaystyle{{{a^2 \epsilon_3}\over{|u|^2}}} & {\rm for} $|u| \ge aR$ 
\boxhwt{25pt}{0pt}{0pt}\cr
}.
\ee
\vskip 7pt
In particular, $\overline\mu_{a,\alpha,\lambda}(u) < 1$ for all $u\in \PA_\infty\HH$ and $\lambda\in\RR$ with $|\lambda|\le\lambda_0$. By upper semi-continuity
of $\overline \mu$, we can extend this estimate to $u$ in a neighbourhood of $\PA_\infty \HH$ and $\lambda$
with $|\lambda|\le\lambda_0$ and $0\le\Im\lambda\le\epsilon$ to conclude
\be{chimubdd}
\chi(u)\,\overline\mu_{a,\alpha,\lambda}(u) \le 1+\epsilon_4,
\ee
where $\epsilon_4$ can be made arbitrarily small by shrinking the support of $\chi$ and taking $\epsilon$ small. 

To estimate the right side of \rf{obv} we consider $u$ in two regions. The first region are the points near
$u\in\RR$ with $|u| \le C$. For these points, the estimate \rf{mubarbdd2} and upper semi-continuity of $\overline\mu_{a,\alpha,\lambda}(u)$ imply
$$
\overline\mu_{a,\alpha,\lambda}(u) \le 1-\epsilon_5
$$
for some $\epsilon_5 >0$. This, combined with \rf{chimubdd}, where we have chosen $\epsilon$ and the support of $\chi$ to make $\epsilon_4$ sufficiently small, proves \rf{mgoal} for these values of $u$.

On the other hand, if $u$ is in the region near $u\in\RR$ with $|u| \ge C$ (including $i\infty$) then $u$ is bounded away from the singularity of $\phi_{\qq_1,\lambda}(u)$ for $\qq_1\in\supp(\nu_a)$, so for these values of $u$ and small $\qq_1$, the values of
$\phi_{\qq_1,\lambda}(u)$ are close to $\phi_{{\bf 0},\lambda}(u)$ and therefore $|\phi_{{\bf \qq_1},\lambda}(u)|$
is uniformly bounded. This means we can exchange the roles of the two factors in \rf{obv} and obtain
\rf{mgoal} for these values of $u$ as well.
\endproof

\vfill\eject